\documentclass[twocolumn, showpacs,preprintnumbers,amsmath,amssymb,prl]{revtex4}

\usepackage{amsmath,amsfonts,amssymb}
 \usepackage{epsf}

\newcommand{\be}{\begin{equation}}
\newcommand{\ee}{\end{equation}}

\newcommand{\bea}{\begin{eqnarray}}
\newcommand{\eea}{\end{eqnarray}}

\newcommand{\bra}{{\langle}}
\newcommand{\ket}{{\rangle}}
\newcommand{\tr}{\hbox{ Tr}}

\begin{document}

\preprint{}

\title{Black Hole Production from High Energy Scattering in AdS/CFT}

\author{Samuel E. V\'azquez}
 \email{svazquez@perimeterinstitute.ca}
 \affiliation{Perimeter Institute for Theoretical Physics, \\  31 Caroline St. North, Waterloo, Ontario, Canada N2L 2Y5. }

\begin{abstract}
In this article we show how to set up initial states in ${\cal N} =4$ SYM theory that correspond to  high energy graviton collisions, leading to black hole formation in $AdS_5\times S^5$. For this purpose, we study states in the gauge theory that are dual to graviton wavepackets localized at the center of $AdS_5$, and  carrying  large angular momentum along the $S^5$. These states are created by exciting only the s-wave mode of one of the complex adjoint scalars of SYM.  For a single graviton, the state is 1/2 BPS and one can show that it is dual to a linearized 1/2 BPS geometry in the bulk. Exploiting this dictionary, we show how to localize the particle's wavefunciton so that the dual linearized metric has the form of a Aichelburg-Sexl shock wave. One can then put two such shock waves into a head-on collision, which is known to produce a trapped surface. Finally, we discuss the prospect of studying graviton scattering directly at strong coupling in the gauge theory using a reduced model of matrix quantum mechanics.

\end{abstract}

\pacs{Valid PACS appear here}

 \maketitle

\section{Introduction}

Since the discovery of the AdS/CFT correspondence \cite{adscft}, there have been great expectations that this duality
will help us answer deep questions about black hole
physics. The reason is that the gauge theory provides a
non-perturbative description of quantum gravity in the
bulk with fixed boundary conditions. In particular, one
expects to find a resolution to the black hole singularity,
and a solution to the information loss problem. For the
former, most research has been focused on finding signatures
of the eternal black hole singularity in the gauge
theory correlation functions \cite{Shenker}. This has been done
exploiting the duality between the gauge theory at finite
temperature, and an eternal AdS-Schwarzchild black
hole in the bulk.

However, one would like to understand the dynamical
phenomena of black hole formation, and its eventual
evaporation.  For this reason, it
is very desirable to understand how to set up initial states in
the gauge theory which are dual black hole forming processes
in the bulk. 
One way to do this  is to study states in ${\cal N} = 4$ SYM
theory which are dual to high energy graviton scattering in the bulk.

 Classically, and in flat space, it is known that for
energies much greater than the Planck's  energy, these
scattering processes  lead to the formation of a trapped surface \cite{Giddings1}. This has been shown by superposing two Aichelburg-
Sexl shock waves, which represent the gravitational field
of massless particles in flat space \cite{AS}. However, quantum mechanically one needs to be more careful and take into account the wave nature of the particle. This was discussed in  \cite{GiddingsWave}.

Scattering processes have also been studied in the context
of the AdS/CFT duality in, for example \cite{scatteringAdSCFT}. However, most
of these works have dealt with the problem of particles
coming from the boundary into the AdS bulk. This has
various problems. The first is that the high redshift between
the bulk and the boundary makes it hard to focus
wavepackets on small scales. Secondly, since these calculation
only involve correlation functions on the boundary,
one is necessarily calculating S-matrix elements. It is
desirable to have a precise state in SYM on $\mathbb{R} \times S^3$,
which we could evolve in time (at least in principle). Finally,
focusing the scattering process into the center of
AdS should lead to simplifications since the whole process
is basically taking place in flat space.

Therefore, in this paper we answer the following simple
question: what is the SYM state dual to a Aichelburg-
Sexl shock wave for a graviton well localized at the center of AdS, and traveling along the $S^5$?  

Moreover, we show
how to superpose two  such waves so we get a two-graviton scattering process in the bulk.  We mainly work at zero impact parameter, but other generalizations are straightforward. 
We also make sure that the resulting
trapped surface is much larger than the string scale, but much smaller that the AdS radius. Furthermore, we ensure that the curvature at the trapped surface is small. Finally, we comment on how one should understand this scattering process at strong coupling and at non-zero impact parameter, in terms of a quantum model of commuting matrices as in \cite{berenstein}.

The organization of the paper goes as follows. In section II, we take a 1/2
BPS linearized perturbation of $AdS_5 \times S^5$ described by
the so-called LLM solution of \cite{LLM}. These solutions can
be classified by ``droplets" on a two dimensional plane.
A droplet with the shape of a unit disk represents the
$AdS_5 \times S^5$ solution. We then make a small ripple along
the boundary of the unit disk. This represents a massless graviton wavefunction. We then take the limit of large radius of $AdS$ while focusing on the region near the wavepacket. This leaves a ten dimensional flat space metric with a linear perturbation of the Aichelburg-Sexl type. The difference here is that we can now control the spread of the wavepacket, and the usual delta function in the metric gets replaced with a smeared finite function. This allow us to control the curvature near the trapped which was an issue discussed in \cite{GiddingsWave}.

In section III, we write down states dual to these LLM shock waves using the dictionary recently developed in \cite{sam}. The states are constructed out of the zero mode of one of the adjoint scalars of SYM theory. We give particular examples of initial states corresponding to two gravitons separated by a finite distance in flat space, and heading towards each other with zero impact parameter. We also comment on the generalization for non-zero impact parameter.

In section IV we speculate on how one should study these initial states and their time evolution, at strong coupling. In particular, we argue that the reduced matrix model of \cite{berenstein} should be sufficient to understand graviton scattering at impact parameters $b > E l_s^2$, where $E$ is the center of mass energy of the collision.

In section V we close with some final comments and future directions. 

{\it Conventions}. In this article we will set the ten dimensional Planck's constant $l_p = 1$. In these units, the $AdS$ radius is related to the rank of the SYM gauge group $N$ by $R = (4 \pi N)^{1/4}$. The string length is given by $l_s = \sqrt{\alpha'} = g^{-1/4}$, where $g$ is the closed string coupling also related to the SYM coupling by $4\pi g = g_{YM}^2$. We will also introduce the parameter $\hbar = 1/N$.

\section{The Aichelburg-Sexl metric from IIB supergravity}
In this section we will show how the Aichelburg-Sexl metric \cite{AS} can be derived from the 1/2 BPS geometries of type IIB supergravity \cite{LLM}.  This metric describes the external gravitational field of a point-like massless particle. In fact, our version of the Aichelburg-Sexl metric will take into account  the wave nature of the particle.
This will allow us to control the curvature of the geometry near the region of the trapped surface.

 All 1/2 BPS solutions to type IIB supergravity with $N$ units of Ramond-Ramond
five-form flux have been found in \cite{LLM}. They preserve a
bosonic $\mathbb{R} \times SO(4) \times SO(4)$ symmetry of the ten
dimensional space-time.  Moreover, they preserve half of the supersymmetries of the original $AdS_5\times S^5$. We will leave the details of this analysis to reference \cite{LLM}. Here we will only need the general form of the metric.

It turns out that all solutions are classified by a single
function, which we  call $\rho$, on a two dimensional plane. The
metrics can be written as, \bea \label{metric}ds^2 &=& -h^{-2}(Dt)^2
 + h^2 (dy^2 + dz d\bar z) + y e^{-G} d\Omega_3^2\nonumber \\
 && +  y e^{G}
 d \widetilde{\Omega}_3^2 , \nonumber \\
 h^{-2} &=& 2 y \cosh G,\nonumber \\
 f &=& \frac{1}{2} \tanh G, \nonumber \\
 \label{f} f(z,\bar z, y) &=&   - \frac{y^2}{2} \int d^2 z' \frac{\rho(z',\bar z')}{(|z -
 z'|^2 + y^2)^2}.
 \eea

 Here, we have defined the covariant derivative, $Dt = dt + V = dt  +
\frac{1}{2} i \bar V dz - \frac{1}{2} i  {V} d\bar z$, and we are
using complex coordinates in the $y = 0$ plane: $z = x_1 + i x_2,
\bar z = x_1 - i x_2$. Moreover, \be \label{V} V(z,\bar z, y)  =
\frac{1}{2} \int d^{2}z' \frac{\rho(z',\bar z')(z - z')}{(|z - z'|^2
+ y^2)^2}\;.\ee

All non-singular solutions must have $\rho = \pm 1/\pi$. Therefore
we can separate the integrations above in domains or ``droplets"
(${\cal D}_i$) for which $\rho = 1/\pi$ (say) inside and $\rho =
-1/\pi$ outside. In this paper we will only consider a single droplet at the center of the complex plane.

The curvature radius of the azymptotic $AdS_5 \times S^5$  ($R$) is set by $R^4  =  \int_{{\cal D}} d^2 z/\pi$. Therefore, we can rescale all {\it spatial}
coordinates by $x^i \rightarrow R^2 x^i$ and we get an overall
factor of $R^2$ in front of the metric. Then, our area
quantization condition is simply,
\be \label{area} \int_{{\cal D}} \frac{d^2z}{\pi}
 = 1\;.\ee
This makes it easier to compare with the gauge theory.
Finally, the energy of the solution can be written as,
\be \label{E} E = J = \frac{1}{\hbar^2 \pi} \int_{{\cal D}} d^2z |z|^2 - \frac{1}{ 2 \hbar^2}\;. \ee

Before turning our attention to the Aichelburg-Sexl metric, lets recover the familiar $AdS_5 \times S^5$ and flat space. For the former, we need to consider a circular droplet of unit radius. The integrations above give,
\be\label{Vads} V = \frac{1 + y^2 + |z|^2 - \sqrt{(1 + y^2 + |z|^2)^2 - 4 |z|^2}}{2 \bar z \sqrt{(1 + y^2 + |z|^2)^2 - 4 |z|^2}}\;.\ee
Similarly,
\be \label{fads} f = \frac{y^2 + |z|^2 - 1}{2 \sqrt{(1 + y^2 + |z|^2)^2 - 4 |z|^2}}\;.\ee
Plugging these results back in the metric, and making the coordinate changes ($z = r e^{i \phi}$),
\bea\label{coordchange} y &=& \sinh \rho \sin \theta  \;, \nonumber \\
r &=& \cosh \rho \cos\theta\;,\\
\tilde \phi &=& \phi - t \;, \nonumber \eea
we obtain the standard form for the $AdS_5\times S^5$ metric:
\bea \label{adsmetric} ds^2 &=&R^2 \left[ -\cosh^2\rho dt^2 + d\rho^2 + \sinh^2\rho d\tilde\Omega_3^2 \right.\nonumber \\
&& \left. + d\theta^2 + \cos^2\theta d\tilde\phi^2 + \sin^2\theta d\Omega_3^2\right]\;.\eea

Let us now obtain flat space. For this, we need to zoom in into the edge of the droplet. We will do it by rescaling
\be \label{scaling} z \rightarrow \left(1 + \frac{x_2}{R^2}\right) \exp\left(\frac{\pi}{2} - \frac{x_1}{R}\right)\;,\;\;y \rightarrow \frac{1}{R^2} y \;,\;\;t \rightarrow \frac{1}{R} t\;,\ee
while taking the limit $R\rightarrow \infty$.  We can then make the following coordinate change: $y = r_1 r_2$, $x_2 = (r_1^2 - r_2^2)/2$. A careful expansion of the metric above gives,
\be \label{flat}ds^2 = - dx_1( 2 dt - dx_1) + d\vec{r}^2\;,\ee
where $\vec{r} = \vec{r_1} + \vec{r_2}$ is an eight dimensional transverse vector to the droplet. Note the difference between this limit and the usual pp-wave limit. In the later, one does not rescale the time coordinate and, instead, make a further rescaling $x_1 \rightarrow x_1/R$.

From the metric (\ref{flat}) we clearly see the two null geodesics at the center of $AdS$: $x_1  = \text{const.}$, and $x_1 = 2 t + \text{const}$. Ripples on the boundary of the droplet following any of these geodesics will be 1/2 BPS, but with opposite angular momentum on the $S^5$. Let us now focus on a ripple centered at $x_1 = 0$. We can later boost this result to obtain a particle along the other geodesic.

The ripple will be given by a fluctuation on the radius of the droplet:\be \label{rpert} r_{\text{boundary}} = 1 +\delta r(\phi)\;.\ee
The area quantization condition (\ref{area}) implies,
\be \label{pertarea}2 \int d\phi \delta r(\phi) + \int d\phi \left[\delta r(\phi)\right]^2 = 0\;.\ee
To leading order in the perturbation, one can set the second term above to zero in most calculations. However, for the energy (\ref{E}) one needs to be careful. Using (\ref{pertarea}) in (\ref{E}) we get,
\be\label{Epert} E \approx \frac{1}{\pi \hbar^2 } \int  d\phi \left[\delta r(\phi)\right]^2\;.\ee

Now we would like to localize the perturbation $\delta r$ within the flat space patch near $\phi = \pi/2$ (or $x_1 = 0$). Moreover, one would like to keep the energy $E$ fixed as we take the $R\rightarrow \infty$ limit.
To this end, we will normalize the solution as
\bea \label{g}\delta r(\phi) &=& \frac{4 \pi^{3/2} \sqrt{E}}{R^3 \sqrt{\lambda}} g\left[R (\phi - \pi/2)/\lambda \right] \nonumber\\
&=& \frac{4 \pi^{3/2}\sqrt{E}}{R^3\sqrt{\lambda}} g(x_1/\lambda)\;.\eea
Here the function $g(x)$ is zero outside the interval $x\in [-1/2,1/2]$, and $\lambda$ is some wavelength that does not scale with $R$. Moreover, the function $g(x)$ is normalized such that,
\be  \int_{-1/2}^{1/2} dx [g(x)]^2 = 1\;.\ee
Using the area constraint (\ref{pertarea}), we also get
\be  \int_{-1/2}^{1/2} dx g(x) = -\frac{2 \pi^{3/2}\sqrt{E}}{R^3\sqrt{\lambda}}  \rightarrow 0\;,\ee
where in the last step we have taken the large $R$ limit.

It is easy to verify from (\ref{Epert}) that this perturbation indeed has energy $E$,  once we take into account the fact that we have rescaled the time as in (\ref{scaling}), and hence the Hamiltonian as $\hat H_{\text{new}} = \hat H_{\text{old}}/R$.

Let us now derive the metric corresponding to this perturbation, starting with the function $f$ of the metric (\ref{f}).  We can write $f  = f_0 + \frac{1}{R^2}\delta f$, where $f_0$ is the AdS space result Eq. (\ref{fads}).
Similarly, $V = V_0 + \delta V$, where $V_0$ is also the AdS space result (\ref{Vads}). The reason for the scalings with R will become apparent in a moment. For now, we just expand the metric to linear order in the perturbations and take the $R\rightarrow \infty$ limit with the change of coordinates (\ref{scaling}). The result is,
\bea ds^2 = - dx_1( 2 dt - dx_1) + d\vec{r}^2 + \delta g_{11} dx_1^2\;,\eea
where
\be \delta g_{11} = i(\delta V - \delta\bar V) - \frac{2 x_2}{y^2} \delta f\;.\ee

Let us now study the form of these perturbations more closely. One can easily show that, after doing the rescalings (\ref{scaling}), the metric perturbation becomes:
\bea \delta g_{11} =  &\approx&  \frac{2 R^3 }{\pi} \int_{-\lambda/2}^{\lambda/2} dx_1' \int_0^{ R^2 \delta r(x_1')}dx_2' \nonumber \\
&&\times \left[x_2' + \frac{1}{2}(x_1^2 - x_1'^2)\right]\nonumber \\
&&\times \left(y^2 +  R^2(x_1 - x_1')^2 + \right.\nonumber \\
 && \left.( x_2 - x_2' - \frac{1}{2}(x_1^2 - x_1'^2))^2 + {\cal O}(1/R)\right)^{-2}\;.\nonumber \\\eea

We can simplify this expression in two ways. First let us note the identity
\bea \lim_{R\rightarrow \infty} \frac{R}{(1 + [R G(x)]^2)^2} &=& -\frac{\partial}{\partial \epsilon} \lim_{R\rightarrow \infty} \left. \frac{R}{\epsilon + [R G(x)]^2} \right|_{\epsilon = 1}\nonumber \\
&=& -\frac{\partial}{\partial \epsilon}   \left. \frac{1}{\sqrt{\epsilon}} \pi \delta(G(x)) \right|_{\epsilon = 1} \nonumber \\
&=& \frac{\pi}{2} \delta\left(G(x)\right)\;.\eea Moreover, from (\ref{g}) we can see that $R^2\delta r\sim 1/R$ and so it is very small for $x_2 \sim {\cal O}(1)$. Therefore, using (\ref{g}) and making the usual change of coordinates $y = r_1 r_2$, $x_2 = (r_1^2 - r_2^2)/2$, we get
\be \delta g_{11} \approx  \frac{(4 \pi)^3 E}{|\vec r|^6} \delta_{\lambda}(x_1)\;,\ee
where we have defined,
\be \delta_\lambda(x)  = \int_{-1/2}^{1/2} dx' [g(x')]^2 \delta(x - \lambda x')\;.\ee
This is basically a delta function with support in an interval $\Delta x = \lambda$.

To summarize, we have found that for a perturbation of the unit disk of the form (\ref{g}), the LLM metric in the large $R$ limit becomes,
\be\boxed{ \label{AS} ds^2 = - dx_1( 2 dt - dx_1) + d\vec{r}^2 + \frac{(4 \pi)^3 E}{|\vec r|^6} \delta_{\lambda}(x_1) dx_1^2\;.} \ee
This is precisely the form of the Aichelburg-Sexl metric in ten dimensions \cite{Giddings1}.
Note that the perturbation obeys the ten dimensional Laplace equation with a delta function source at $\vec r = 0$.  This delta function is only an artifact of the limit $R \rightarrow \infty$. The full ten dimensional solution is completely smooth even at $\vec r = 0$. 

We can then see that for a 1/2 BPS perturbation with the opposite angular momentum one can simply shift this result by $x_1 \rightarrow 2 t - x_1$. At the level of linearized gravity, we can superimpose both waves and get a solution to Einstein's equations. Of course, the solution will not be BPS anymore. Moreover, the simple superposition is only valid outside the region $x_1, 2t - x_1 > 0$, where the colliding shocks start to influence each other.  The full solution inside this region is still not known.

 The two shock waves will collide at $t = 0$ and at $x_1 = 0$.  It is known that such head-on collisions lead to the formation of an eight dimensional closed marginally  trapped surface with radius \cite{Giddings1}
\be \label{rh}|\vec r_h| \sim E^{1/7}\;.\ee

We now need to ensure that this classical calculation is not spoiled by high curvatures near the trapped surface. This issue was discussed in \cite{GiddingsWave}. For a single shock wave, the only non-zero components of the curvature tensor are \cite{curvature}
\be R_{x_1 i x_1 j} = -\frac{1}{2} \delta_\lambda(x_1) \frac{\partial^2}{\partial r_i \partial r_j}   \frac{(4 \pi)^3 E}{|\vec r|^6}\;.\ee
All curvature invariants are finite unless at the precise location of the particle $\vec r = 0$.

For the two-wave superposition, one encounters a singularity at $t = 0, x_1 = 0$ if we take the limit $\lambda \rightarrow 0$. In fact, in this limit the only diverging curvature invariant is of the form \cite{GiddingsWave}
\be (R_{\mu \nu \rho \sigma})^2 \sim \frac{E^2}{|\vec{r}|^{16} \lambda^2}\;,\ee where we have taken into account the peak in the function $\delta_\lambda(x)$. We can evaluate this quantity at the radius of the trapped surface (\ref{rh}), and require: $(R_{\mu \nu \rho \sigma})^2 \ll 1/l_s^4$, $|\vec{r}_h| \gg \lambda$ and $|\vec r_h| \gg l_s$.  These conditions translate into

\begin{itemize}
  \item $ g^{-1/2} E^{-1/7} \ll \lambda \ll E^{1/7}$,
  \item $E^{1/7} \gg g^{-1/4}$.
\end{itemize}
It is clear that for large energies these constraints are very easy to satisfy and they allow us to take $\lambda \ll 1$.

At this point, we would like to emphasize that in order to obtain the flat space limit with a fixed string length, we cannot take the usual 't-Hooft limit. That is, we do not take $g_{YM}^2 N  = \text{fixed}$. Instead, we take $N \rightarrow \infty$ keeping $g_s$ small but fixed. Indeed, this is basically the same limit one takes when studying pp-waves \cite{BMN}.

%%%%%%%%%%%%%%%%%%%%%%%%%%%%%%%%%%%%%%%%%%%%%%%%%%%%%%%%%%%%%%%%%%%%%%

\section{Initial States from ${\cal N}  = 4$ SYM}

In this section we will review the basic dictionary between the 1/2 BPS states of ${\cal N}  = 4$ SYM theory on $\mathbb{R}\times S^3$, and matrix eigenvalue distributions. These eigenvalue distributions can then be mapped to the LLM geometries studied in the previous section \cite{LLM}. Using this map it is easy to find the dual state to the Aichelburg-Sexl metric (\ref{AS}). Moreover, it is straightforward to write a superposition of two shock waves.

The 1/2 BPS states of ${\cal N} = 4$ SYM theory are created with the zero mode of one of the complex adjoin scalar field $Z(t,\Omega)$, where $\Omega \in S^3$. The tree-level quadratic action for this mode is simply
\be\label{Zmodes} S =
 \int dt \tr \left(|\dot Z|^2 -
|Z|^2\right)\;.\ee 
This simple model has been discussed in \cite{12BPS}.

We can now introduce the usual oscillators.
We start with the following
canonical commutation relations, \be [ (Z)_i^j, (\Pi)_k^l] = i
\delta_i^l \delta_k^j \;,\;\;\;\; [ (\bar Z)_i^j, (\bar \Pi)_k^l] =
i \delta_i^l \delta_k^j\;,\ee where $\Pi= \dot{\bar
Z}$, and $\bar \Pi= \dot{Z}$. The harmonic oscillator operators are defined as usual, \be A^\dagger =
\frac{1}{\sqrt{2}} \left( Z- i{\bar \Pi} \right)\;,\;\;\;\;\;\;  \bar A^\dagger =
\frac{1}{\sqrt{2}}  \left( \bar Z -i
\Pi \right)\;, \ee with \be [A_i^j, (A^\dagger)_k^l] =
 \delta_i^l \delta_k^j\;,\;\;\;\;  [\bar A_i^j,
(\bar A^\dagger)_k^l] = \delta_i^l \delta_k^j\;,\ee
Note that $\Pi^\dagger = \bar \Pi$.

The Hamiltonian of the system is simply \footnote{We are ignoring the constant from normal ordering which gets canceled with the fermions in the full supersymmetric theory.},
\be \label{H} H = \tr\left( A^\dagger A+ \bar{A}^\dagger \bar{A}\right)\;.\ee
Note that these operators carry a single unit of the $U(1) \subset SU(4)$ R-charge. The generator for this $U(1)$ is
\be J = \tr\left( A^\dagger A- \bar{A}^\dagger \bar{A}\right)\;.\ee
So we see that $A^\dagger$ has charge $+1$ and $\bar A^\dagger$ has charge $-1$. This R-charge is dual to an angular momentum along the $S^5$ in the bulk.
Since $[H, J] = 0$, one can define a new Hamiltonian,
\be \label{Hprime}H' = H - J\;,\ee
which is naturally identified with the time $t$ in the LLM coordinates used in the previous section.

The (anti) 1/2 BPS states  are formed by acting with only ($\bar A^\dagger$) $A^\dagger$  on the ground state. In the time slicing (\ref{Hprime}), the 1/2 BPS states are time independent and the anti-1/2 BPS have a simple time dependence where $\bar A^\dagger \rightarrow e^{ 2 i t} \bar A^\dagger$.

Lets start by studying ``coherent states" of the form,
\be \label{BPSstate}|\psi_{\text{1/2 BPS}} \ket \propto e^{\tr \Omega(A^\dagger)}|0\ket\;.\ee
We can also build non-BPS states that include two wavefunctions with opposite R-charge:
\be\label{scattstate} |\psi \ket \propto e^{\tr \Omega_1(\bar A^\dagger)} e^{\tr \Omega_2(A^\dagger)}|0\ket\;.\ee
In the dual string theory, these states will correspond to two gravitons at the center of $AdS_5$ and traveling in opposite directions along the equator of $S^5$.

Although the state (\ref{scattstate}) is not BPS, as long as the gravitons are initially well separated in the $S^5$, one can treat each exponential as a separate initial 1/2 BPS state. This will allow us to use the dictionary between the 1/2 BPS states (\ref{BPSstate}) and the LLM geometries developed in \cite{sam}. Let us now review this dictionary.

We can start by defining coherent states for the annihilation operator $A$. Namely, $A_i^j |Z\ket = Z_i^j|Z\ket$ and where the resolution of the identity takes the form $1 = \int [d^2 Z] |Z\ket\bra Z|$,
with $[d^2 Z]$ the $U(N)$ invariant measure over complex matrices.

The 1/2 BPS state (\ref{BPSstate}) becomes,
\be \bra Z| { \psi}\ket
= e^{\tr \; \Omega(Z)/\hbar} e^{-\tr|Z|^2/(2\hbar)}\equiv \psi(Z)e^{-\tr|Z|^2/(2\hbar)} \;,\ee
where we have rescaled the $Z$ matrix for later convenience.
By going to an eigenvalue basis, one can show that the normalization of this state
is given by the partition function \cite{sam}, \be \label{part}\bra { \psi} | { \psi}\ket
\propto \prod_{i = 1}^N \int d^2 z_i e^{\sum_j W(z_j,\bar{
z}_j)/\hbar + \sum_{i < j} \log |z_i - z_j|^2 }\;,\ee where \be
\label{W} W(z, \bar z) = - |z|^2 + \Omega(z) + \overline{
\Omega(z)}\;.\ee

It is well known that in the ``classical" limit $\hbar \rightarrow
0$, one can use the saddle point approximation and replace the sums in (\ref{part}) by integrals over an eigenvalue density in the complex plane \cite{matrixreview}. This is the usual
2D Coulomb gas problem. 

The ``free energy" functional is,
\bea F[\rho] &=&  -\frac{1}{\hbar} \int d^2 z \rho(z) W(z,\bar z) \nonumber \\
&&- \frac{1}{2} \int d^2z \int d^2 z' \rho(z) \rho(z') \log|z - z'|^2 \;.\eea The saddle point equations are $\delta F[\rho] = 0$ subject to the constraint $\int d^2 z \rho = \hbar^{-1}$. Using the fact that $\log |z|^2$ is the Green's function in two dimensions,  one finds that the density $\rho$ is locally constant with
$\rho = 1/(\hbar \pi)$. The eigenvalues are then distributed over these constant density droplets which can be  translated to the LLM geometries. Compelling evidence for this was specially given in \cite{sam,diego}, where it was shown that one can recover a probe string sigma model on a LLM geometry in the gauge theory by including a single trace ``probe" state in top of the 1/2 BPS state (\ref{BPSstate}).

The energy of a 1/2 BPS state can be calculated by the free field theory Hamiltonian (\ref{H}). Using the coherent states, the Hamiltonian is simply $H = \tr(Z\partial Z)$, and it acts on the wavefunction $\psi(Z)$. Doing a partial integration in the Random Matrix Model, it is easy to show that the energy of the state is given by,
\be E = \frac{1}{\hbar} \bra \tr|Z|^2\ket  - \frac{1}{2\hbar^2} \approx  \frac{1}{\hbar^2\pi} \int_{\cal D} d^2 z |z|^2 - \frac{1}{2\hbar^2}\;.\ee This expression matches precisely with the energy calculated in the gravity side, Eq.  (\ref{E}).

We can now take a potential of the form
\be \Omega(z) = \sum_{k >0} t_k z^k\;.\ee
It is known that for a single droplet the moments $t_k$ are related to the shape of the droplet by \cite{matrixreview}
\be\label{exactmoments} t_k = \frac{1}{k} \oint_{\partial {\cal D}} \frac{dz}{2\pi i} \bar z z^{-k}\;.\ee
For a linearized perturbation in the boundary of the form (\ref{rpert}), the moments reduce to,
\be \label{moments} t_k \approx \frac{1}{\pi k}\int_0^{2\pi} d\phi \delta r(\phi) e^{-ik\phi}\;.\ee

Therefore, we can find the gauge theory dual to the Aichelburg-Sexl metric (\ref{AS}). We simply use (\ref{g}) in (\ref{moments}):
\be \label{ASmoments} t_k \approx \frac{\hbar}{k} \sqrt{\frac{E \lambda}{\pi}} e^{-i k \pi/2} \int_{-1/2}^{1/2} dx g(x) e^{i k x \lambda /R}\;.\ee
The single particle state can now be written as,
\bea\label{finalBPSstate} |\psi\ket &=& \exp\left[ - \sqrt{\frac{E \lambda}{\pi}}  \right. \nonumber \\
&& \left.  \times \int_{-1/2}^{1/2} dx g(x) \tr \log\left(i e^{- i  x \lambda /R} - \sqrt{\hbar} A^\dagger \right)\right] |0\ket \;. \nonumber \\
\eea
It is now very easy to write down the anti-BPS state which is located at the null geodesic $x_1 = L$ at $t = 0$. We simply shift $x \rightarrow x  + L$ and replace $A^\dagger \rightarrow \bar A^\dagger$ inside the log of (\ref{finalBPSstate}).  The two particle initial state is simply the product of two such functions acting on the ground state as in (\ref{scattstate}).

\subsection{The Problem with Single Trace States}
In this section we would like to point out an obstruction in constructing the Aichelburg-Sexl metric using only single trace states in SYM.

 Consider the 1/2 BPS state
\be \label{state}|\psi\ket  = \sum_{n >0 } \frac{\psi_n \hbar^{n/2}}{\sqrt{n}} \tr\left(A^\dagger\right)^n |0\ket\;.\ee   In the large $N$ limit, this state has unit norm so long as,
\be  \label{normwave}\sum_{n >0} |\psi_n|^2 \equiv 1\;.\ee
Of course, this is only valid as long as the amplitudes $|\psi_n|^2$ decay sufficiently fast for $n > n_\text{max.}$, where $n_\text{max} \ll N$. This constraint comes from the fact that, for sufficiently large $n$, the single grace states with different $n$s are no longer orthogonal.

The ``free energy" functional that comes from computing the square amplitude using coherent states is
\bea\label{freeenergy} F[\rho] &=&  \frac{1}{\hbar} \int d^2 z \rho(z) |z|^2 \nonumber \\
&&- \frac{1}{2} \int d^2z \int d^2 z' \rho(z) \rho(z') \log|z - z'|^2 \nonumber \\
&&- \log \left| \int d^2 z\rho(z) \left( \sum_{n >0 } \frac{\psi_n }{\sqrt{n}}  z^n\right)\right|^2  \;.  \nonumber \\
\eea 
The saddle point equation $\delta F[\rho] = 0$ with the constraint $\int d^2 z \rho = \hbar^{-1}$ gives,
\bea \text{const.} &=&  -\frac{1}{\hbar}  \left(-|z|^2+ \sum_{k >0} t_k z^k + h.c.\right)\nonumber\\
&& - \int d^2 z' \rho(z') \log|z - z'|^2\;, \eea
where the constant comes from the Lagrange multiplier implementing the area constraint, and the moments $t_k$ are given by,
\bea \label{tpsirel}t_k &=&    \frac{\hbar  \psi_k}{\sqrt{k} \left(   \int d^2 z \rho(z) \sum_{n >0} \frac{ \psi_n}{\sqrt{n}} z^n\right)} \nonumber \\
&=&  - \frac{\hbar^2  \psi_k}{\sqrt{k} \left(  \sum_{n >0} \sqrt{n} \psi_n t_{-n} \right)}\;.\eea
In the last line, we have used the definition of the moments in terms of the eigenvalue distribution, Eq. (\ref{exactmoments}).  Moreover, at the linearized level, one can see from (\ref{moments}) that $t_{-n} \approx - t_n^*$. 

One can then use (\ref{tpsirel}) and the normalization (\ref{normwave}) to solve for the wavefunction:
\bea\label{tpsirel2} \psi_k &=& \hbar^{-1}\sqrt{k} t_k\;,\\
\label{hbarnorm}\hbar^2 &=& \sum_{n>0} k |t_k|^2  \;.\eea
The last equality comes from the normalization of the wavefunction, and gives an extra constraint on the moments of the distribution. Note that we did not have this constraint with the exponential states. This new constraint is part of the problem. To see this, lets construct a particular example.

 Consider a perturbation $\delta r(\phi) $ of the form (\ref{g}) with \be g(x) \propto \left\{ 
\begin{array}{cc}
  1&   x\in [-1/2,0) \\ 
  -1 &  x\in [0,1/2]\\
  0 & \text{otherwise}
\end{array}
\right. \;.\ee 
Using the linearized form of the moments (\ref{ASmoments}) one can easily finds,
\be\label{example} \psi_k = -\frac{4 i \hbar R}{k^{3/2}}\sqrt{\frac{E}{\pi \lambda}} e^{-i k\pi/2} \sin^2\left(\frac{k\lambda}{4 R}\right)\;.\ee
The normalization of the wavefunction gives a relation between $E$ and $\lambda$:
\be \label{rel} E \approx \frac{\pi}{\lambda \log 2 }\;.\ee
To obtain this relation, one approximates the sum over $k$ in (\ref{hbarnorm}) by an integral.

The relation (\ref{rel}) is not present for the exponential states studied above.  Moreover, we see that this is not compatible with the constraints studied in the previous section. In particular, we want to take $E \gg 1$ to have a macroscopic trapped surface. This means that we need $\lambda \ll 1$.  In this case, however,  the low curvature conditions would imply $\lambda^{6/7} \gg g^{-1/2}$ which is not compatible with $l_s/ l_p \sim g^{-1/4} \gg 1$. Therefore, we conclude that single trace states are not good enough to produce macroscopic black holes.

There is another problem with single trace states: the ten dimensional wavefunction on the bulk is not well localized in the orthogonal directions to the shock wave. We show this in detail in the Appendix, by relating the single trace excitations to single particle states in the bulk. However, one would like to understand this problem directly from the gauge theory. So far we haven't found a clear answer to this question. What seems to be happening is that the saddle point approximation is not good when we use the single trace states. That is, due to the logarimic term in the ``free energy"  (\ref{freeenergy}), the eigenvalue distribution has a lot of variance around the saddle point. It would be interesting to show this by a numerical simulation as in \cite{davidcotta}.

Nevertheless, we know that this does not happen for the exponential states (\ref{BPSstate}).   Indeed, we know how to compute the corrections to the saddle point in this case. This is discussed in \cite{matrixreview}. The leading correction is a smoothing of the density distribution $\rho(z)$ near the boundary of the droplet. That is, the density is not really a step function, but it has an exponential decay at the edge of the droplet over a distance of order $\sim \sqrt{\hbar}$  (in units where the circular droplet has unit radius). This correction is even present for the ground state circular droplet.  If we translate this to the flat space coordinates (\ref{scaling}), this means that the variation of the density occurs on scales of $x_2 \sim {\cal O}(1)$.

Why don't we see this effect in String Theory? The reason is that,  the string scale is very big  $l_s = g^{-1/4}$ compared to $l_p \equiv 1$, and so the string sees a step function to a very good accuracy.  Therefore, the supergravity solution captures only the average shape of the droplet.  This applies to general droplets. In other words, we only care about the shape of the droplet and not its small scale structure along the edge. The shape of the droplet will be accurately captured by the saddle point approximation as long as the variation of the shape of the edge occurs on a scale $\Delta\phi \gg \sqrt\hbar$. That is, we need a ``long" wavelength variation. 
 In our case, we saw in the previous section that $\Delta \phi \sim 1/R \sim \hbar^{1/4} \gg \sqrt \hbar$. Therefore, we are allowed to use the saddle point approximation for the exponential states.

\section{The Strong Coupling Description}

In the previous section, we have seen how to construct initial states in SYM theory which are dual to high energy graviton collisions on the $S^5$. However, this has been done using only the weak field theory description. For the case of states with the same R-charge, this description is enough since they are supersymmetric. However, when it comes scattering processes, one is dealing with non-BPS states and therefore one needs to work directly at strong coupling.  In this section we argue that graviton scattering at non-zero impact parameter should be described by a reduced quantum mechanical model of commuting matrices along the lines of \cite{berenstein}.

Lets begin by reviewing the proposal of \cite{berenstein}.   One starts by studying the dynamics of the zero modes of the six hermitean scalars of  SYM, $X^a$ ($a = 1,2,\ldots, 6$). The classical hamiltonian for these modes is given by \cite{davidcotta},
\bea
\label{Heff}
H_{eff} &=&  \tr\left( \sum_{a=1}^6 \frac 12 (D_t X^a)^2 + \frac 1 2 (X^a)^2 \right. \nonumber \\
&& \left. +\sum_{a,b=1}^6\frac{g_{YM}^2}{8\pi^2} [X^a,X^b][X^a,X^b]\right)
\eea

On general grounds, one expects the size of the SYM ground state to be,
\be\label{groundsize} \frac{1}{N}\sum_{a = 1}^6\bra 0| \tr(X^a X^a)|0\ket \sim N\;, \ee
in the large $N$ limit.  Therefore, if we re-scale the scalar by $X^a \rightarrow \sqrt{N} X^a$,  we see that the commutator term in (\ref{Heff}) will have an extra $g_{YM}^2 N$ factor in front of it. It then makes sense that, in the large $N$ limit (or in the large 't-Hooft limit), the lowest energy fluctuations around the ground state should be described by a quantum mechanical system of {\it commuting} matrices. 

For commuting matrices we can go to an eigenvalue basis. In this basis we can define a vector containing the eigenvalues of all the matrices, $\vec{x}_i = (X_{ii}^1,\ldots, X_{ii}^6)$,
where $i   = 1, \ldots, N$.
 The quantum Hamiltonian that follows from (\ref{Heff}) in this basis becomes \cite{berenstein},
\be H = \sum_i \left(-\frac{1}{2\mu^2} \nabla_i \mu^2 \nabla_i + \frac{1}{2} |\vec{x}_i|^2\right)\;,\ee
where
\be \mu^2 = \prod_{i < j} |\vec{x}_i - \vec{x}_j|^2\;,\ee
comes from the Jacobian produced from going to the eigenvalue basis.

The wavefunctions must be symmetric under the exchange of the eigenvalues by gauge invariance. Moreover, the measure in this basis becomes,
\be \bra \psi |\psi\ket = \int \prod_i d^6 x_i \mu^2 \psi^*\psi\;.\ee
We can get rid of the extra factor of $\mu^2$ in the inner product by rescaling the wavefunction as $\psi \rightarrow \psi/\mu$. After this rescaling, the Hamiltonian becomes,
\bea \label{eigenH} H &=& \sum_i\left( -\frac{1}{2\mu} \nabla_i \mu^2 \nabla_i \frac{1}{\mu} + \frac{1}{2} |\vec{x}_i|^2\right) \nonumber \\
&=& \sum_i\left(- \frac{1}{2}\nabla_i^2 + \frac{1}{2} |\vec{x}_i|^2 \right)+ V_\text{eff}\;,\eea
where,
\bea V_\text{eff} &=&  \sum_i \frac{1}{2} \frac{1}{\mu} \left( \nabla_i^2 \mu \right)^2 \nonumber \\
&=& - 6 \sum_{i\neq j} \frac{1}{|\vec{x}_i - \vec{x}_j|^2}\nonumber \\
&& + \sum_i \sum_{j,k \neq i} \frac{(\vec{x}_i - \vec{x}_j)\cdot(\vec{x}_i - \vec{x}_k)}{|\vec{x}_i - \vec{x}_j|^2 |\vec{x}_i - \vec{x}_k|^2}\;.\eea

The ground state of this Hamiltonian is exactly known and takes the form \cite{berenstein}
\be \label{groundstate}\psi_0 \sim \mu \exp\left(-\frac{1}{2} \sum_i |\vec{x}_i|^2\right)\;.\ee
In the large $N$ limit, this wavefunction leads to an eigenvalue distribution which consist of an $S^5 \subset \mathbb{R}^6$.  The radius of the sphere was calculated in \cite{bvc} and it is given by $r_0 = \sqrt{N/2}$. This confirms the expectation (\ref{groundsize}).

Now we want to add the dynamics of the off-diagonal elements of the matrices, for both the zero modes, and the higher spherical harmonics on $S^3$.  For this we follow the prescription of \cite{berenstein, bvc, berenscorrea,japs,menew}, and quantize the off-diagonal modes in the {\it background} of the diagonal eigenvalue distribution. More specifically, one can derive an effective quadratic Hamiltonian for these modes that looks like
\be \label{offdiagH} H_\text{off diag.} \sim \sum_{i\neq j} \sum_{\alpha} w^{(\alpha)}_{i,j} (A_\alpha^\dagger)_i^j (A_\alpha)_j^i \;,\ee
with \be\label{disp} w^{(\alpha)}_{i,j} = \sqrt{m_\alpha^2 + \frac{g_{YM}^2}{2\pi^2} |\vec{x}_i - \vec{x}_j|^2}\;.\ee 
The operators $A_\alpha, A_\alpha^\dagger$ obey the usual commutator relations for harmonic oscillators. The collective index $\alpha$ goes over the different kinds of excitations (scalars of fermions), and their spherical harmonic numbers.  Furthermore,  $m_\alpha$ is their tree level mass, which is a positive (half) integer for (fermions) scalars. The coupling to the eigenvalues comes from the commutator terms in the SYM lagrangian.

Of course, with no supersymmetry one expects that integrating out the off-diagonal modes will produce a non-trivial functional determinant that will badly modify the Hamiltonian of the eigenvalues. The assumption of \cite{berenstein} is that, given all the supersymmetry of SYM theory, the functional determinants between bosons and fermions should roughly cancel to leading order. It would be interesting to prove this rigorously.

Now, the proposal of \cite{bvc, berenscorrea, japs} is that we should treat the $\vec{x}_i$ as random numbers given by the distribution of the ground state (\ref{groundstate}). Given the size of he ground state $r_0  = \sqrt{N/2}$, one can see from (\ref{offdiagH}) that the off-diagonal modes will generically have masses of the order $m_\text{off-diag.} \sim \sqrt{g_{YM}^2 N}$, and will decouple in the large 't-Hooft limit, or large $N$ limit.

 It was first found in \cite{bvc} that, using this approach, one could reproduce the energy of the so-called BMN states to all loops in the 't-Hooft coupling. Moreover, a picture of semi-clasical ``string bits" emerged which was later on confirmed in the classical string theory by  the so-called ``Giant Magnons" of Maldacena and Hofman \cite{magnons}. 
 
 Now, let us suppose that we perturb the ground state by creating two waves of the form,
\be \psi \sim e^{\sum_i \Omega_1(\vec{x}_i)  + \sum_j \Omega_2 (\vec{x}_j) }\psi_0\;.\ee
Let us also suppose that each individual wave $\Omega_a$ is ``almost" BPS. By this we mean that, if we had only one of them, the state would be very close to a BPS state. For example, one can show \cite{davidcotta} that holomorphic functions like $\Omega \sim \sum_k t_k z^k$ are approximately BPS in the sense that $\bra H - J\ket \approx 0$ in the large $N$ limit.  

Finally, lets suppose that the two waves create two density fluctuations in the eigenvalues $\delta \rho_a(\vec x)$ that are both localized within a ``flat space" patch.  If we normalize the radius of the $S^5$ eigenvalue distribution to one, we have seen in section II that the flat space patch is located within an angular difference  of $\Delta \phi \sim 1/R \sim N^{-1/4}$. The two density fluctuations will be located within this patch. That is, their angular separation is $\sim b/N^{1/4}$, where $b$ is a {\it fixed} number.

We can now ask: are there any light off-diagonal modes that can be excited? One can divide the off-diagonal excitations into two categories. First, there are the ``string bits" that connect different eigenvalues within the same density fluctuation $\delta \rho_a$. These should be suppressed since we assumed that each individual wave is almost BPS. However, we also have the string bits that join eigenvalues between the two different lumps. From the dispersion relation (\ref{disp}), it is easy to see that these off-diagonal modes will have energies of the order
\bea E_{\text{off-diag}} &\sim& \sqrt{m_\alpha^2 + \frac{g_{YM}^2 N}{2\pi^2} \left(\frac{b}{N^{1/4}}\right)^2 } > \sqrt{g} b R\;. \nonumber \\
\eea
However, we need to remember, that we are looking at small time scales and that we re-defined our time coordinate by $t \rightarrow t/R$, and so the Hamiltonian by $H \rightarrow H R$. Therefore, the energy of the off-diagonal modes in the new time coordinate is $E_{\text{off-diag}} > \sqrt{g} b =  l_s^{-2} b$.  To be able to ignore these modes, one  would need to have large impact parameters so that $E < E_{\text{off-diag}}$, where $E$ is the center of mass energy of the collision. This leads to the bound,
\be\label{bbound} b > l_s^2 E\;.\ee

In the string theory picture, this is precisely the bound to create long strings \cite{Gross}.  It is interesting to reproduce this bound from a gauge theory perspective.

However, to create black holes one needs $b \sim |\vec r_h| \sim E^{1/7}$, which is much smaller than (\ref{bbound}).  This seems to suggest that the matrix model can only capture graviton scattering in the Born approximation. Nevetheless, there is evidence that stringy physics does not play a role in the formation of the black hole \cite{Gross,nonlocality}. It might be that the matrix model is good even for impact parameters lower than (\ref{bbound}). In any case, it would be interesting to even recover the $1/r^2$ gravitational potential from the matrix model.

Now let us explain in a bit more detail, how one should attack these problems. First, of all we need to learn how to write down well localized initial states using the $SO(6)$ matrix model. The guess would be that one can take the same $\Omega$ that we used for the single matrix model in section III. This needs to be checked in more detail. In particular, one would like to see if the resulting density fluctuation (for a single graviton) is indeed localized as desired. This might involve the need of numerical simulations as in \cite{davidcotta}.

To control the impact parameter using our formalism, one can start with an holomorphic wave of the form $\Omega \sim \sum_k t_k z^k$ (and the corresponding anti-holomorphic one), and perform a rotation of the coordinate $z$.  This will take one of the geodesics off the equator. Then, by controlling the initial angle from which we start our lumps, one can arrange an initial state with two lumps that, after time evolving, will miss each other by a finite impact parameter. 

Finally, one needs to solve the time evolution of the density lumps $\delta \rho_a$ using the Hamiltonian (\ref{eigenH}).  Note that this is a non-trivial $N$-body problem. However, one expects some simplifications in taking the large $N$ limit.

\section{conclusion}

In this paper we have made a first attempt to carefully define initial states in SYM theory that, under time evolution, are dual to high energy graviton scattering in the bulk.  The basic approach has been to concentrate on the zero modes of the SYM scalars, which are dual to excitations that are located at the center of $AdS_5$, but carrying angular momentum along the $S^5$. We have given a dictionary between gauge theory states and linearized LLM geometries. Furthermore, we have shown how to obtain the Aichelburg-Sexl metric in the bulk, and how to superpose two such shock wave to produce a {\it classical} trapped surface. 

We have also paid special attention to suppressing  stringy effects in the bulk, by producing a smoothed-out version of the shock-wave geometry.  Finally, we have given evidence that scattering of gravitons a finite impact parameter is described by a reduced model of matrix quantum mechanics. 

There are many interesting question one can try to answer along these lines. First, one would like to learn how to construct the bulk wavefunction dual to the exponential BPS states (\ref{BPSstate}). For single trace states, the dictionary is discussed in the Appendix. However, for these states, we showed that the wavefunction is not well localized in the directions orthogonal  to the propagation of the particle.  We argued that this is not a problem for the exponential states based on the validity of the saddle point approximation in the Random Matrix model. Nevertheless, it is very desirable to have an actual wavefunction in the bulk to make an explicit calculation.

Another interesting avenue of research is the study of gravitons scattering using the reduced matrix model studied in the previous section.  In particular one would like to understand how to time evolve density fluctuations $\delta \rho$ in the eigenvalue distribution. According to the arguments given above, we expect that such scattering processes will capture the physics of graviton scattering at large impact parameter in flat space. It would be very interesting to confirm this.

The reduced matrix model could also help to answer the question about the wavefunction localization. For example, one can study the a holomorphic wavepacket of the form $\exp( \sum_i \sum_k t_k z_i^k)$ and use the full $SO(6)$ eigenvalue measure to study its spread along all the directions on the $S^5$. This will require the use of numerical techniques such as in \cite{davidcotta}.

It would be very interesting to understand how to calculate the finite time propagator $\bra \psi_1(t)|\psi_2(0)\ket$ in the matrix model. In fact,  we expect that for short times $t \sim 1/R$, and for density fluctuations localized within an angular distance $\Delta \phi \sim 1/R$, this computation will give us the flat space S-matrix. One can then start to look for signatures of the black hole formation. In particular, one expects that these amplitudes should behave like 
\be \bra \psi_1(t)|\psi_2(0)\ket \sim e^{-S_{B H}} \sim e^{- E^{8/7}}\;,\ee
 for large energies $E$ \cite{Giddingstalk}. Finally, note that the small black hole  has a finite entropy $S_{BH} \sim E^{8/7}$ in the large $N$ limit. Therefore, the matrix model (even without the off-diagonal excitations) has more than enough degrees of freedom to account for its microstates (I thank Rob Myers for pointing this out).

\section{acknowledgments}
 I would like to thank Steve Giddings, David Berenstein, Rob Myers and the String Theory group at the Perimeter Institute for many interesting discussions.  This work was supported in part by the Department of Energy grant DOE-FG02-91ER40618. Research at Perimeter Institute for Theoretical Physics is supported in part by 
the Government of Canada through NSERC and by the Province of Ontario through 
MRI.

 \section{appendix: the ten dimensional wavefunction}

In this appendix we study the relation between the single trace wavefunction (\ref{state}) in SYM, and the ten dimensional wavefunction in  the bulk. The main conclusion is that the single trace states lead to wavepackets that are not well localized in the direction orthogonal  to the propagation of the graviton.

 To find the ten dimensional wavefunction, we need to linearize the LLM solutions, and quantize the fluctuation around $AdS_5\times S^5$. Fortunately, this was done in \cite{marsano}.  Their parametrization of the edge fluctuation $\delta r(\phi)$ takes the form,
 \be \label{fluc} \delta r(\phi) = \sum_{n\neq 0} a_n e^{i n \phi}\;,\ee
 where $a_{-n} = a_n^*$.

 After quantization, one finds that the Fourier coefficients $a_n$ obey the commutation relations (after proper normalization),
 \be \label{commutators} [a_n, a^\dagger_m] = n \delta_{n,m}\;.\ee
 This is very similar to the expectation value,
 \be \bra 0|\hbar^{n/2} \tr A^n \hbar^{m/2}\tr(A^\dagger)^m|0\ket  = n \delta_{n,m}\;.\ee
 Therefore we can identify,
 \be\label{stateid} a_n^\dagger |0\ket \sim \tr (A^\dagger)^n |0\ket\;.\ee
 
 The metric fluctuation in global coordinates looks like \cite{marsano},
 \be \delta g \sim \sum_{n\neq 0} c_n Y_n s_n\;,\ee
 where $c_n$ are constants, and
 \be \label{modes}Y_n =  e^{ i n (\phi - t)} \cos^{|n|}\theta\;,\;\;\; s_n \propto\frac{1}{\cosh^{|n|}\rho} a_n e^{i n t}\;.\ee
 The important point here is that these modes satisfy the ten dimensional Laplace equation:
 \be \nabla_{10}^2 (s_n Y_n) = 0\;.\ee
 Therefore, they provide a position representation of the wavefunction corresponding to the mode $a^\dagger_n|0\ket$. In other words,
 \be \bra \vec x_{10}| a^\dagger_n |0\ket \sim s_n Y_n\;.\ee
 Using the identification (\ref{stateid}), we see that the modes $s_n, Y_n$ also give ten dimensional interpretation of the single trace states in the gauge theory.
 
 Let us now define the normalized modes,
 \be \phi_n(\vec x) \propto s_n Y_n\;,\ee
 so that
 \be \int_\Sigma \phi^*_n \phi_m = \delta_{n,m}\;.\ee
 Here, $\Sigma$ is a constant time surface.  More explicitly,
 \be \phi_n \propto \frac{\sqrt{(n+1)(n+2)(2n-1)(2n-3)} \cos^n \theta }{\cosh^n \rho} e^{i n \tilde \phi}\;.\ee

There is a subtlety with the $n = 1$ mode which is not normalizable. In  the LLM picture, this mode corresponds to a translation of the droplet in in the complex plane. We can set this mode to zero by translating the droplet. For example, shifting the whole droplet by a complex number $a$, the moments change by
 \be t_k \rightarrow t_k + a^* \delta_{k,1} - a (k+1) t_{k+1} + {\cal O}(a^2)\;.\ee
 Since the moments that we are considering are very small, we can just set $a^* = -t_1$ and get rid of this mode. The other modes will not be changed to leading order.

 A general normalized wavefunction is,
 \be \psi(\vec x) = \sum_{n >1} \psi_n \phi_n(\vec x)\;,\ee
 with
 \be \sum_{n >1}|\psi_n|^2 = 1\;.\ee
 Given the relation between the bulk and gauge theory states (\ref{stateid}), it is natural to identify $\psi_n$ above with the corresponding coefficients used in the single trace gauge theory state (\ref{state}).  Let us now use the example studied in the previous section. 

To study the spread of the wavefunction, is is convenient to make some coordinate transformations. 
Using the coordinate change (\ref{coordchange}), the rescalings (\ref{scaling}) and the relation $\cos \theta = r$ one can easily show,
\be \frac{\cos\theta}{\cosh \rho} \approx 1 - \frac{\vec{r}^2}{2 R^2}\;,\ee
where $\vec r$ is the eight-dimensional vector transverse to the propagation of the graviton. Therefore, the probability density in a spatial slice $\Sigma$ is,
\bea\label{fullwave} |\psi(\vec{x})|^2 &\sim& |\vec r|^7 \left|\sum_{k>1} \psi_k \sqrt{(n+1)(n+2)(2n-1)(2n-3)} \right. \nonumber \\
&& \left.  \times \left(1 - \frac{\vec{r}^2}{2 R^2}\right)^n e^{i n \tilde \phi} \right|^2\;.\eea 
Here, we have included the measure for the eight transverse directions. 

We can now look at the example studied in section III. Namely, the wavefunction given by Eq.  (\ref{example}). After integrating out the $\tilde \phi$ coordinate, one finds that the full wavefunction   (\ref{fullwave}) is peaked around $|\vec r| \sim \sqrt{R}$, with a spread of similar size.  This is way too big! Therefore, we conclude that this kind of wavefunction cannot localize the graviton within the flat space region.


\begin{thebibliography}{99}

\bibitem{adscft}
  J.~M.~Maldacena,
  ``The large N limit of superconformal field theories and supergravity,''
  Adv.\ Theor.\ Math.\ Phys.\  {\bf 2} (1998) 231
  [Int.\ J.\ Theor.\ Phys.\  {\bf 38} (1999) 1113]
  [arXiv:hep-th/9711200].
  %%CITATION = HEP-TH 9711200;%%
  %\cite{Witten:1998qj}

  E.~Witten,
  ``Anti-de Sitter space and holography,''
  Adv.\ Theor.\ Math.\ Phys.\  {\bf 2}, 253 (1998)
  [arXiv:hep-th/9802150].
  %%CITATION = HEP-TH 9802150;%%

  S.~S.~Gubser, I.~R.~Klebanov and A.~M.~Polyakov,
  ``Gauge theory correlators from non-critical string theory,''
  Phys.\ Lett.\ B {\bf 428}, 105 (1998)
  [arXiv:hep-th/9802109].
  %%CITATION = HEP-TH 9802109;%%


\bibitem{Shenker}
  P.~Kraus, H.~Ooguri and S.~Shenker,
  ``Inside the horizon with AdS/CFT,''
  Phys.\ Rev.\  D {\bf 67}, 124022 (2003)
  [arXiv:hep-th/0212277].


  L.~Fidkowski, V.~Hubeny, M.~Kleban and S.~Shenker,
  ``The black hole singularity in AdS/CFT,''
  JHEP {\bf 0402}, 014 (2004)
  [arXiv:hep-th/0306170].

\bibitem{Giddings1}
  D.~M.~Eardley and S.~B.~Giddings,
 ``Classical black hole production in high-energy collisions,''
  Phys.\ Rev.\  D {\bf 66}, 044011 (2002)
  [arXiv:gr-qc/0201034].

\bibitem{AS}
P. C. Aichelburg and R. U. Sexl,
``On the gravitational field of a massless particle," Gen. Rel. Grav. {\bf 2}, 303 (1971).

\bibitem{linearpert}
  L.~Grant, L.~Maoz, J.~Marsano, K.~Papadodimas and V.~S.~Rychkov,
  ``Minisuperspace quantization of 'bubbling AdS' and free fermion  droplets,''
  JHEP {\bf 0508}, 025 (2005)
  [arXiv:hep-th/0505079].


\bibitem{GiddingsWave}
  S.~B.~Giddings and V.~S.~Rychkov,
  ``Black holes from colliding wavepackets,''
  Phys.\ Rev.\  D {\bf 70}, 104026 (2004)
  [arXiv:hep-th/0409131].

V.~S.~Rychkov,
  ``Black hole production in particle collisions and higher curvature
  gravity,''
  Phys.\ Rev.\  D {\bf 70}, 044003 (2004)
  [arXiv:hep-ph/0401116].



\bibitem{scatteringAdSCFT}
 J.~Polchinski,
  ``S-matrices from AdS spacetime,''
  arXiv:hep-th/9901076.
 
  S.~B.~Giddings,
  ``Flat-space scattering and bulk locality in the AdS/CFT  correspondence,''
  Phys.\ Rev.\  D {\bf 61}, 106008 (2000)
  [arXiv:hep-th/9907129].
 
 S.~B.~Giddings,
  ``The boundary S-matrix and the AdS to CFT dictionary,''
  Phys.\ Rev.\ Lett.\  {\bf 83}, 2707 (1999)
  [arXiv:hep-th/9903048].

  A.~Jevicki and H.~Nastase,
  ``Towards S matrices on flat space and pp waves from SYM,''
  arXiv:hep-th/0501013.

\bibitem{berenstein}
  D.~Berenstein,
  ``Large N BPS states and emergent quantum gravity,''
  JHEP {\bf 0601}, 125 (2006)
  [arXiv:hep-th/0507203].

 \bibitem{LLM}
  H.~Lin, O.~Lunin and J.~M.~Maldacena,
  ``Bubbling AdS space and 1/2 BPS geometries,''
  JHEP {\bf 0410}, 025 (2004)
  [arXiv:hep-th/0409174].
  
  

\bibitem{sam}
  S.~E.~Vazquez,
  ``Reconstructing 1/2 BPS space-time metrics from matrix models and spin
  chains,''
  Phys.\ Rev.\  D {\bf 75}, 125012 (2007)
  [arXiv:hep-th/0612014].

\bibitem{curvature}
  V.~S.~Rychkov,
  ``Black hole production in particle collisions and higher curvature
  gravity,''
  Phys.\ Rev.\  D {\bf 70}, 044003 (2004)
  [arXiv:hep-ph/0401116].

\bibitem{BMN}
  D.~Berenstein, J.~M.~Maldacena and H.~Nastase,
  ``Strings in flat space and pp waves from N = 4 super Yang Mills,''
  JHEP {\bf 0204}, 013 (2002)
  [arXiv:hep-th/0202021].


\bibitem{12BPS}
S.~E.~Vazquez,
  ``BPS condensates, matrix models and emergent string theory,''
  JHEP {\bf 0701}, 101 (2007)
  [arXiv:hep-th/0607204].

  D.~Berenstein,
  ``A toy model for the AdS/CFT correspondence,''
  JHEP {\bf 0407}, 018 (2004)
  [arXiv:hep-th/0403110].

 D.~Yamada,
  ``Quantum mechanics of lowest Landau level derived from N = 4 SYM with
  chemical potential,''
  arXiv:hep-th/0509215.
 
 T.~Harmark and M.~Orselli,
  ``Quantum mechanical sectors in thermal N = 4 super Yang-Mills on R x
  S**3,''
  Nucl.\ Phys.\  B {\bf 757}, 117 (2006)
  [arXiv:hep-th/0605234].
 
  Y.~Takayama and A.~Tsuchiya,
  ``Complex matrix model and fermion phase space for bubbling AdS
  geometries,''
  JHEP {\bf 0510}, 004 (2005)
  [arXiv:hep-th/0507070].
  
  A.~Ghodsi, A.~E.~Mosaffa, O.~Saremi and M.~M.~Sheikh-Jabbari,
  ``LLL vs. LLM: Half BPS sector of N = 4 SYM equals to quantum Hall  system,''
  Nucl.\ Phys.\  B {\bf 729}, 467 (2005)
  [arXiv:hep-th/0505129].
  
  A.~Donos, A.~Jevicki and J.~P.~Rodrigues,
  ``Matrix model maps in AdS/CFT,''
  Phys.\ Rev.\  D {\bf 72}, 125009 (2005)
  [arXiv:hep-th/0507124].
  
  
 
 
  
  \bibitem{diego}
  H.~Y.~Chen, D.~H.~Correa and G.~A.~Silva,
  ``Geometry and topology of bubble solutions from gauge theory,''
  Phys.\ Rev.\  D {\bf 76}, 026003 (2007)
  [arXiv:hep-th/0703068].


\bibitem{matrixreview}
  A.~Zabrodin,
  ``Matrix models and growth processes: From viscous flows to the quantum  Hall
  effect,''
  arXiv:hep-th/0412219.






\bibitem{davidcotta}
  D.~Berenstein and R.~Cotta,
  ``A Monte-Carlo study of the AdS/CFT correspondence: An exploration of
  quantum gravity effects,''
  JHEP {\bf 0704}, 071 (2007)
  [arXiv:hep-th/0702090].

\bibitem{bvc}
  D.~Berenstein, D.~H.~Correa and S.~E.~Vazquez,
  ``All loop BMN state energies from matrices,''
  JHEP {\bf 0602}, 048 (2006)
  [arXiv:hep-th/0509015].

\bibitem{berenscorrea}
  D.~Berenstein and D.~H.~Correa,
  ``Emergent geometry from q-deformations of N = 4 super Yang-Mills,''
  JHEP {\bf 0608}, 006 (2006)
  [arXiv:hep-th/0511104].

\bibitem{japs}
  Y.~Hatsuda and K.~Okamura,
  ``Emergent classical strings from matrix model,''
  JHEP {\bf 0703}, 077 (2007)
  [arXiv:hep-th/0612269].
  
  \bibitem{menew}
  D.~Berenstein and S.~E.~Vazquez,
  ``Giant magnon bound states from strongly coupled N=4 SYM,''
  arXiv:0707.4669 [hep-th].

\bibitem{magnons}
  D.~M.~Hofman and J.~M.~Maldacena,
  ``Giant magnons,''
  J.\ Phys.\ A  {\bf 39}, 13095 (2006)
  [arXiv:hep-th/0604135].

\bibitem{Gross}
  S.~B.~Giddings, D.~J.~Gross and A.~Maharana,
  ``Gravitational effects in ultrahigh-energy string scattering,''
  arXiv:0705.1816 [hep-th].
  

\bibitem{nonlocality}
  S.~B.~Giddings,
  ``Locality in quantum gravity and string theory,''
  Phys.\ Rev.\  D {\bf 74}, 106006 (2006)
  [arXiv:hep-th/0604072].
  %%CITATION = PHRVA,D74,106006;%%

\bibitem{Giddingstalk}
Talk given at ``Santa Barbara Gravity Workshop 2007", URL: http://www.physics.ucsb.edu/~giddings/sbgw/.

\bibitem{marsano}
  L.~Grant, L.~Maoz, J.~Marsano, K.~Papadodimas and V.~S.~Rychkov,
  ``Minisuperspace quantization of 'bubbling AdS' and free fermion  droplets,''
  JHEP {\bf 0508}, 025 (2005)
  [arXiv:hep-th/0505079].


\end{thebibliography}
\end{document}